\documentclass[]{rsos}

\usepackage[super,sort&compress,comma]{natbib}

\begin{document}

\title{Dust evolution, a global view: \\ I. Nano-particles, nascence, nitrogen and natural selection \ldots joining the dots}

\author{
A. P. Jones}  

\address{Institut d'Astrophysique Spatiale, CNRS, Univ. Paris-Sud, Universit\'e Paris-Saclay, B\^at. 121, 91405 Orsay cedex, France}

\subject{Astrochemistry, Astrophysics}


\corres{A.P. Jones\\
\email{Anthony.Jones@ias.u-psud.fr}}

\begin{abstract}
The role and importance of nano-particles for interstellar chemistry and beyond is explored within the framework of {\em The Heterogeneous dust Evolution Model at the IaS} (THEMIS), focussing on their active surface chemistry (nascence), the effects of nitrogen doping and the natural selection of interesting nano-particle sub-structures. 
Nano-particle driven chemistry, and in particular the role of intrinsic epoxide-type structures could provide a viable route to the observed gas phase OH in tenuous interstellar clouds en route to becoming molecular clouds. 
The aromatic-rich moieties present in asphaltenes likely provide a viable model for the structures present within aromatic-rich interstellar carbonaceous grains. 
The observed doping of such nano-particle structures with nitrogen, if also prevalent in interstellar dust, is likely to have important and observable consequences for surface chemistry and the formation of pre-cursor pre-biotic species. 
\end{abstract}

\maketitle

\section{Introduction}
\label{sect_intro}

The smaller end of the interstellar dust size distribution, which extends from $\mu$m down to sub-nm sizes, is at the heart of many important astrophysical processes, such as; the origin of the IR emission bands, the heating of the gas through photo-electric emission and the catalytic formation of molecular hydrogen on grain surfaces. 
The latter being the major driving force for most interstellar chemistry. An understanding of the peculiarities of nano-particle physics within the interstellar medium (ISM) is therefore absolutely essential but has not yet been fully recognised or explored. 
In most dust studies the physical properties of nano-particles are simply treated as an extrapolation of those of bulk materials extended down to nm sizes. 
Such an approach is over-simplistic because the size-dependent properties of interstellar analogue materials can help us to self-consistently understand and explain many of the critical observable properties of dust in the ISM.\cite{2012A&A...542A..98J,2013A&A...558A..62J} 
Here some of the consequences of adopting a more realistic approach to the chemistry and physics of nano-particles are explored. 

The paper is organised as follows: 
Section~\ref{sect_THEMIS} briefly describes the fundamentals of the assumed dust model (THEMIS), 
Section~\ref{sect_nps} focusses on the nature of interstellar nano-particles, 
Section~\ref{sect_nascence} considers the active chemistry of nano-particle surfaces (nascence),  
Section~\ref{sect_N} investigates the role of nitrogen atom doping in carbonaceous nano-grains, 
Section~\ref{sect_natsel} speculates on the nature and natural selection of possibly pre-biotic aromatic-type moieties and 
Section~\ref{sect_conclusions} concludes this work.

\section{THEMIS}
\label{sect_THEMIS}

The ideas presented here and in the accompanying papers\cite{ANT_RSOS_topdown,ANT_RSOS_globules} arise from a re-consideration of the nature of interstellar dust within the framework of {\em The Heterogeneous dust Evolution Model at the IaS} (THEMIS).\cite{2013A&A...558A..62J,2014A&A...565L...9K,2016A&A...999A..99J} 
This new dust modelling approach makes the fundamental assumption that the interstellar silicate and carbonaceous dust populations, both essentially completely amorphous in nature, cannot be completely segregated in the ISM because of the cycling of matter, especially carbon, between the gas and dust phases. This re-cycling is due to destructive processing in energetic environments such as shocks and photo-dissociation regions with intense stellar UV/EUVradiation fields. 
The result is that amorphous silicate grains are mixed with a carbonaceous dust component, most likely in core/mantle (CM) structures.\cite{1986Ap&SS.128...17G,1989ApJ...341..808M,1990QJRAS..31..567J,2013A&A...558A..62J} 
The foundation of the THEMIS dust modelling approach is its diffuse ISM dust model,\citep{2013A&A...558A..62J,2014A&A...565L...9K} which comprises the following grain and core/mantle (CM) grain structures and compositions: 
\begin{itemize}
\item  a steep power-law distribution of hydrogen-poor amorphous carbon, a-C, nano-particles (radii $0.4 \leqslant a \lesssim 20$\,nm) with strongly size-dependent optical properties and most of the mass in the smallest nano-particles, {\it i.e.} $dn(a)/da \propto a^{-5}$,  
\item  a log. normal distribution ($a \simeq 10-3000$\,nm, $a_{\rm peak} \simeq 160$\,nm) of large hydrogenated amorphous carbon, a-C(:H), CM grains with UV photolysed a-C surface layers (mantle depth $\simeq 20$\,nm) surrounding hydrogen-rich amorphous carbon cores, a-C:H, and 
\item  a log. normal distribution ($a \simeq 10-3000$\,nm, $a_{\rm peak} \simeq 140$\,nm) of large amorphous silicate grains, a-Sil, of olivine-type and pyroxene-type composition with Fe metal and FeS nano-paricle inclusions, a-Sil$_{\rm Fe,FeS}$, with a-C mantles (depth $\lesssim 10$\,nm) formed by carbon accretion and/or by the coagulation of small a-C particles onto their surfaces. 
\end{itemize}
This diffuse ISM dust model is consistent with the interstellar IR-UV extinction, IR emission bands, FIR-mm thermal emission and dust scattering (albedo) and the observed variations of these properties in the diffuse ISM and also in the transition from diffuse to dense regions.\cite{2013A&A...558A..62J,2014A&A...565L...9K,2015A&A...579A..15K,2015A&A...577A.110Y,2015A&A...580A.136F,2015A&A...000A.000J,2015A&A...000A.000Y}

\section{Nano-particles}
\label{sect_nps}

A recent re-assessment of the composition and nature of interstellar carbonaceous nano-particles resulted from a calculation of their chemical, structural and optical properties, {\it i.e.}, a derivation of the size and surface dependence of their complex refractive indices\citep{2012A&A...540A...1J,2012A&A...540A...2J,2012A&A...542A..98J} and the use of these within a new evolutionary interstellar dust model.\cite{2013A&A...558A..62J} Perhaps one of the major conclusions from these works is that the interstellar nano-particle observables, {\it i.e.}, FUV extinction, the UV bump, the IR emission bands and the MIR emission continuum, that had previously been attributed to different populations, are actually the various manifestations of the size-dependent properties of a single population of hydrogenated amorphous carbon, a-C(:H), nano-particles (radii 0.4 to $\sim 20$\,nm). Here the term a-C(:H) is used to encompass the entire family of aliphatic-rich, H-rich (a-C:H) to aromatic-rich, H-poor (a-C) hydrogenated amorphous carbon semi-conducting materials. \cite{1986AdPhy..35..317R,2000PhRvB..6114095F,2004PhilTransRSocLondA..362.2477F} In fact, the above-mentioned evolutionary dust model\cite{2013A&A...558A..62J} requires a large fraction ($\approx 60\%$) of the carbonaceous material dust mass to be in particles with radii less than 20\,nm. Thus, a-C(:H) nano-particles would appear to be a particularly important constituent of interstellar dust. 

The nature of a-C(:H) nano-particles and their evolution in circumstellar regions was explored within the context of fullerene formation in planetary nebul\ae.\cite{2012ApJ...757...41B,2012ApJ...761...35M} The structure of these particles at nm size scales is rather interesting because they are an intimate and amorphous mix of aliphatic, olefinic and aromatic carbon structures.\cite{2012A&A...542A..98J,2012ApJ...761...35M} The smallest of them, the so-called "arophatic" particles,\cite{2012ApJ...761...35M} contain perhaps only a hundred or so carbon atoms and consist of small aromatic domains containing only a few rings ($N_{\rm R} = 1$ to 3) that are linked into a contiguous network by relatively short chains ($\approx 4$ C atoms in length) of bridging aliphatics and/or olefinics.\citep{2012A&A...540A...1J,2012A&A...540A...2J,2012A&A...542A..98J,2015A&A...581A..92J}  The consequences of the chemical, structural and physical properties of these structures is explored in the rest of this paper.

\subsection{Asphaltenes as a guiding framework} 
\label{sect_asphaltenes}

Building upon the recently-presented ideas on the origin of the diffuse interstellar bands\cite{ANT_RSOS_topdown} it appears that poly-heterocyclic structures within interstellar a-C(:H) nano-particles could play a key role in their chemistry within the ISM. This work\cite{ANT_RSOS_topdown,ANT_RSOS_globules} profits, in no small part, from the detailed analyses of the asphaltene moieties (extracted  from petroleum and coal), which were studied using atomic force microscopy, molecular orbital imaging and scanning tunnelling microscopy.\citep{JACS_2015_Asphaltenes} These asphaltene moieties were shown to exhibit a wide range of complex aromatic structures, with no two having the same structure, and often with methyl and large alkyl peripheral side-groups. While six-fold aromatic rings are predominant in the analysed species, they also contain a significant fraction of five-fold rings and rare seven-fold rings. Typically, asphaltene structures consist of a single aromatic core with alkyl side-groups, which may be up to $\simeq 2$\,nm ($\simeq 15$ C atoms) long.\citep{JACS_2015_Asphaltenes} The asphaltene morphologies are described and analysed in more detail in the following paper in this series.\cite{ANT_RSOS_topdown}

\subsection{The kangaroo's tail and the jelly bowl scenario} 
\label{sect_jelly}

The stability of nano-particles, and more generally of nano-structures, against the destructive effects of FUV photon absorption in the tenuous ISM is a matter that has perhaps not previously been given sufficient consideration. For example, carbonaceous nano-particles will almost certainly be significantly dehydrogenated by FUV photon-driven photolysis in the diffuse interstellar medium\cite{1996A&A...305..616A} but they can also dissociate as a result of the excitation arising from energetic (FUV) photon absorption.\cite{1996A&A...305..602A,2010A&A...510A..36M,2010A&A...510A..37M} Nevertheless, and whatever the final outcome of the excitation, they will be significantly vibrationally excited by the absorption of these energetic photons and can re-configure their structure as a result of this excitation,\cite{2012ApJ...761...35M} a well-known effect in irradiated materials.\cite{Banhart_RepProgPhys:1999p1181} 
Thus, any channels that can evacuate excess energy from a nano-particle or nano-structure can only increase its stability. 
In this sense, it may be that the aliphatic/olefinic bridges and/or tails within the contiguous network of an a-C(:H) nano-particle, and similarly the alkyl side-groups on asphaltene moieties, can perhaps give an extra degree of stability to these "arophatic" nano-species (as a kangaroo's tail does to a kangaroo) by providing an  absorber to dampen the destructive effects of large energy fluctuations. 
Increased internal energies (due to absorbed photons) will likely promote the "floppy" aliphatic/olefinic bridges/tails into more highly vibrationally-excited states than the more "rigid" aromatic domains, which require higher internal energies to excite them. 
The preferential vibrational excitation of the aliphatics/olefinics would allow the particle to shed energy via IR emission from these states. 
At high enough internal energies the particle will have few energy loss options other than dissociation {\it i.e.}, by fragmenting.\cite{2010A&A...510A..36M,2015A&A...581A..92J}
This is a bit like a glass bowl full of jelly; below some threshold tapping or shaking to bowl only makes the jelly wobble but shaking it above threshold deposits too much energy into the system and the glass bowl fragments spraying jelly blobs all over the place ({\it e.g.}, when the bowl of jelly is dropped!).

\section{Nascence}
\label{sect_nascence}

Nascence in nano-particles is an enhanced surface reactivity effect, which can lead to the formation of interesting chemical structures and surface reaction products. 
In the low-density ISM, nano-particles carry a substantial fraction of the total grain surface area available for surface chemistry and are therefore expected to play an important role in interstellar chemistry. 
In most interstellar chemistry studies the grain surfaces are treated as passive substrates providing sites for reactions between adsorbed species or between adsorbed and incident gas phase species, which may be atoms, ions, radicals or molecules. 
This is the classical scenario for molecular hydrogen formation on grains surfaces in the ISM. 
However, a-C(:H) nano-particle surfaces are chemically (re)active\cite{2015A&A...581A..92J} and therefore able to interact with incident species in ways that have not yet been considered in interstellar chemistry studies, particularly in the diffuse/translucent ISM interfaces at the onset of molecular cloud formation. 
Thus, nascent interstellar nano-particles could be an important driver of interstellar chemistry in transition regions where they are abundant, the gas density is high enough for a significant flux of gas phase species onto their surfaces and the interstellar UV radiation field can also play a role. 
For instance, nano-particle surfaces will preferentially react with abundant gas phase species such as H, O and N atoms and Mg$^+$, Si$^+$, Fe$^+$ and S$^+$ ions in diffuse/translucent regions.
Given that H atoms are the most abundant lightweights in the diffuse atomic ISM we consider them separately and then consider the surface chemistry with the heavyweights.

\subsection{H atom interactions and FUV photolysis}
\label{sect_H_nascence}

The evolution of carbonaceous dust in the more tenuous regions of the ISM is likely driven by H atom interactions leading to CH bond formation, which is counterbalanced by CH bond dissociation by far-ultraviolet (FUV) photolysis.\cite{2001A&A...367..347M,2001A&A...367..355M,2008ApJ...682L.101M,2010ApJ...718..867M,2012A&A...542A..98J,2014A&A...569A.119A,2015A&A...581A..92J,2015A&A...584A.123A} 
In the diffuse ISM the carbonaceous nano-particles must be aromatic-rich (H-poor) in order to explain the $3-13\,\mu$m IR emission bands and the dust thermal emission at MIR-mm wavelengths.\cite{2012A&A...540A...2J,2012A&A...542A..98J,2013A&A...558A..62J,Faraday_Disc_paper_2014}  If we make the assumption that interstellar carbonaceous nano-particles are optically thin to photon absorption at FUV wavelengths\cite{2012A&A...542A..98J} and are collisionally thin to interactions with H atom incident from the gas, {\it i.e.},   the H atoms can interact with any available adsorption sites that might be available to form CH bands without hinderance. 
Then the key parameter is the ratio of the CH bond FUV photo-dissociation rate to the CH bond formation rate by H atom addition. The latter rate is given by 
\begin{align}\label{eq_H_int}
R_{\rm H} \ \simeq \ n_{\rm H} \ \sigma_f \ v_{\rm H}, 
\end{align}
where $n_{\rm H}$ is the gas phase H atom density in the tenuous ISM (typically $\simeq 30$ H atoms\,cm$^{-3}$), 
$\sigma_f$ is the CH bond formation cross-section ($\simeq 2 \times 10^{-18}$ cm$^2$)\cite{2010ApJ...718..867M} and $v_{\rm H}$ is the H atom speed ($v_{\rm H} = [8 k_{\rm B} T_{\rm kin}/\{\pi m_{\rm H}\}]^{0.5} = 1.3 \times 10^5$ cm\,s$^{-1}$ at $T_{\rm kin} = 80$\,K $\equiv 0.01$\,eV). 
The counteracting CH bond FUV photo-dissociation rate is  
\begin{align}\label{eq_UV_int}
R_{\rm diss.} \ \simeq \ F_\lambda \ e^{-\tau_\lambda} \ \sigma_{\rm CH},  
\end{align}
where $F_\lambda$ is the FUV photon flux ($3 \times 10^7$ photons\,cm$^{-2}$\,s$^{-1}$)\cite{2002ApJ...570..697H}, $e^{-\tau_\lambda}$ is the attenuation of the FUV radiation field at optical depth $\tau_\lambda$ (hereafter assumed to be zero for the tenuous and optically-thin diffuse ISM) and $\sigma_{\rm CH}$ is the CH bond FUV photo-dissociation cross-section ($\simeq 10^{-19}$\,cm$^{-2}$).\cite{1972JChPh..57..286W,1994CPL...227..243G,2001A&A...367..355M,2001A&A...367..347M,2014A&A...569A.119A} 
The critical ratio of these two rates, $\zeta$, is then 
\begin{align}\label{eq_H_UV_rat}  
\zeta \ = \ \frac{R_{\rm diss.}}{R_{\rm H}} \ = \ \frac{F_\lambda \ \sigma_{\rm CH} }{n_{\rm H} \ \sigma_f \ v_{\rm H}}.  
\end{align}
If we substitute the above-cited values into Eq.(\ref{eq_H_UV_rat}) we find $\zeta \simeq 0.4$, seemingly slightly favouring hydrogenation over photolysis.  However, it should be noted that while $\sigma_{\rm CH}$ has been measured for methane molecules,\cite{1994CPL...227..243G} which is probably a good approximation for the CH bonds in a-C(:H) nano-particles, the value of $\sigma_f$, the CH bond formation cross-section, was obtained for bulk materials rather than nano-particles. Given these uncertainties, particularly for nano-particles, and the uncertainties in the interstellar FUV photon flux, there is seemingly a rather fine balance between FUV photolysis and re-hydrogenation in the more tenuous ISM. 
Further, while all a-C(:H) nano-particles likely contain photolysable CH bonds they may not always express suitable sites for H atom addition, {\it i.e.}, the efficiency for (re-)hydrogenation may be intrinsically lower than that for CH bond photolysis. 
Given that the observational evidence points to the dominance of aromatic-rich (nano-)particles to explain the dust IR-mm emission features and continuum in the tenuous ISM, where there is little extinction, it would appear that FUV photolysis must be favoured over the effects of hydrogenation\cite{2012A&A...542A..98J,2013A&A...558A..62J,2012ApJ...760L..35L} but that moderate extinction will likely tip the balance in favour of hydrogenation.\cite{Faraday_Disc_paper_2014}

\subsection{Interaction with the heavies}
\label{sect_heavy_nascence}

Given that heavy atoms (O, C, N, \ldots) are about four orders of magnitude less abundant than H atoms in the gas in the ISM, it is clear that for them $\zeta \gg 1$ (see Eq.~\ref{eq_H_UV_rat}) and so, during their surface residence time, an adsorbed/chemisorbed heavy atom X will be subject to a significant number of UV photon absorption events that will affect its surface residence time and adsorbed state, and could drive the catalytic (dissociative) formation of small radicals. 
The intrinsic structure of an a-C(:H) nano-particles includes domains with aliphatic, olefinic and aromatic C$-$H and C$-$C, C=C bonding, of which the unsaturated olefinic, and to a lesser extent the aromatic, C=C bonds will be receptive to adsorptive chemical bond formation with incident gas phase atoms (O, C, N, \ldots). 
In the diffuse ISM reactions with the neutral, O and N atoms and C$^+$ ions will be favoured because of their high gas phase abundances. Further, and given that the smallest but most abundant nano-particles will be predominantly neutral,\cite{2001ApJS..134..263W} reactions with the   less abundant ions Mg$^+$, Si$^+$, Fe$^+$ and S$^+$ will not be hindered.

\begin{figure}[!h]
\centering\includegraphics[width=4.0in]{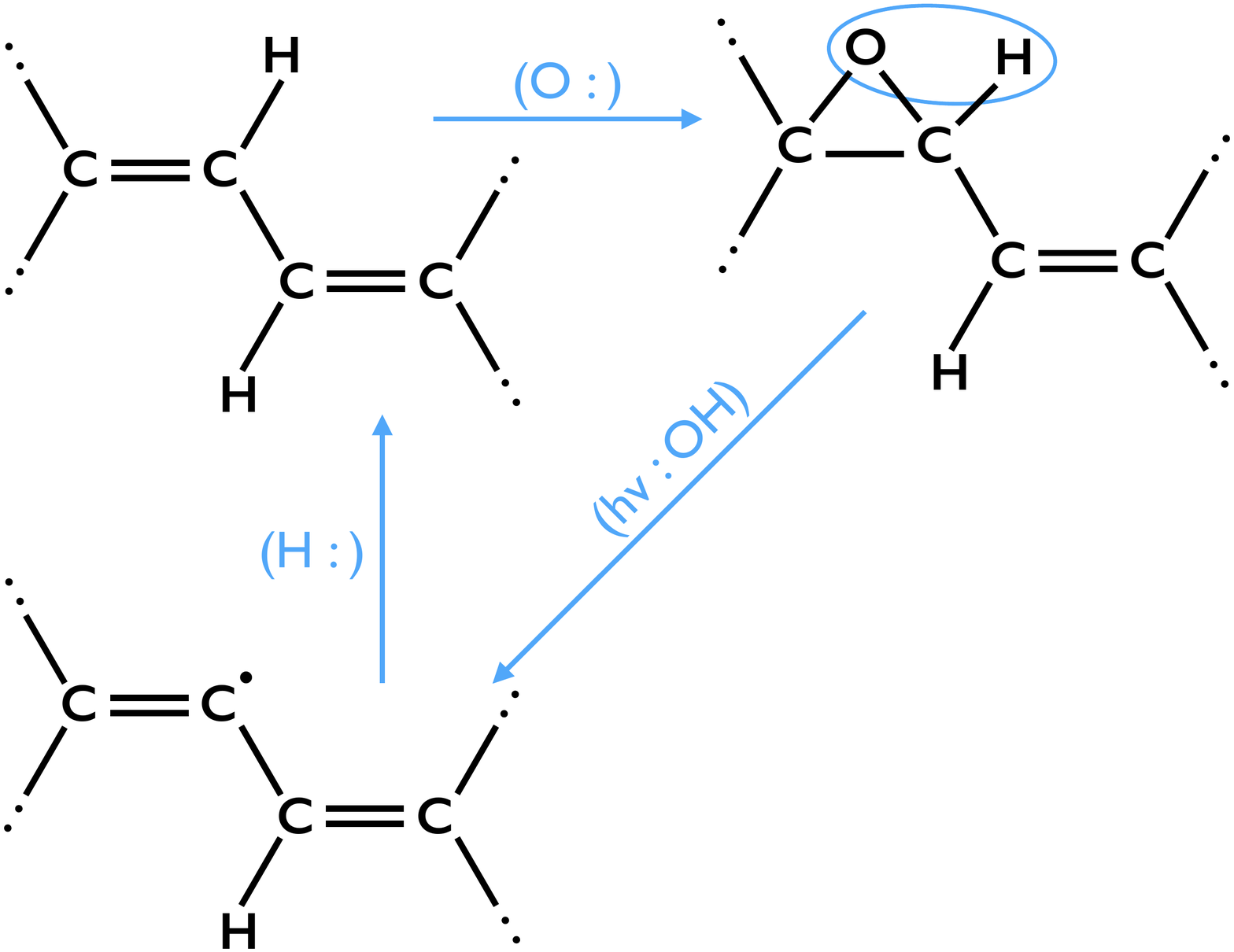}
\caption{A formation pathway for nano-particle surface epoxide groups and the catalytic formation and liberation of OH radicals following UV photolysis in the tenuous ISM.}
\label{fig_epoxide_OH}
\end{figure}

\begin{figure}[!h]
\centering\includegraphics[width=4.0in]{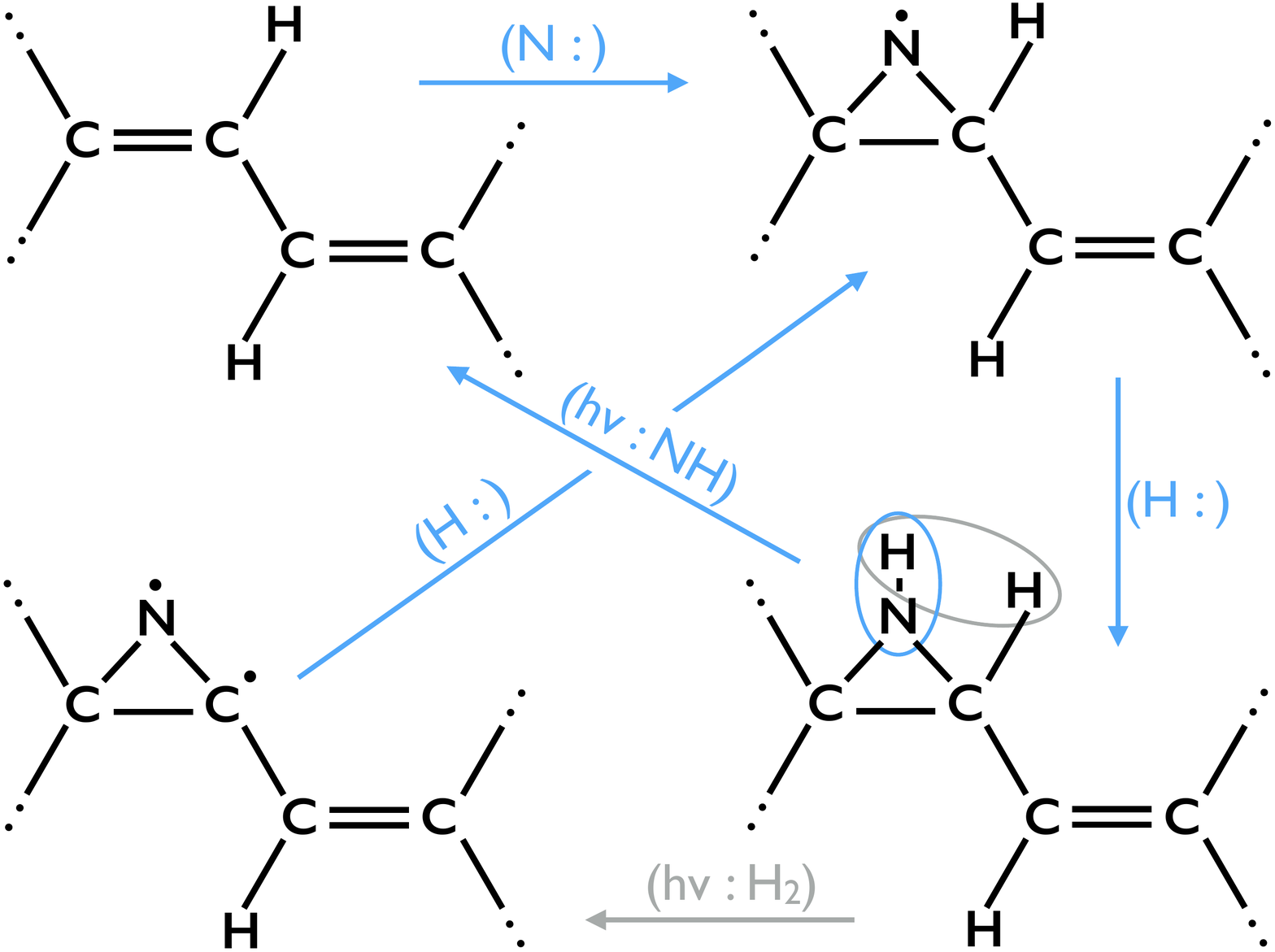}
\caption{A formation pathway for nano-particle surface aziridine groups and the catalytic formation and the liberation of H$_2$ molecules and NH radicals following UV photolysis in the tenuous ISM.}
\label{fig_aziridine_NH}
\end{figure}

For example, it is highly probable that O and N atoms will react with the olefinic double bonds, $>$C=C$<$, that link the aromatic domains, rather than with the more stable aromatic C=C bonds, in a-C(:H) nano-particles.\cite{2015A&A...581A..92J} 
These reactions will form epoxide-type groups with O atoms\cite{2011ApJ...741..121W} and aziridine-type groups with N atoms, the latter being the nitrogen analogue of the epoxide group. In each of these chemical species the fundamental structure consists of a C$_2$X three membered ring $>$C$_-^{\rm X}$C$<$ (X = O or N), which is very reactive and readily decomposes upon UV photolysis. Thus, the epoxide and aziridine groups could be key intermediates in the photon-driven formation of small radicals in tenuous interstellar clouds.  
In such regions the most likely products of the stellar UV photolysis of epoxide- and aziridine-type groups would be OH and NH via the following reaction pathways, where the X indicates either O or N atoms: \\[0.2cm] 
\hspace*{0.03cm} $^{\rm H}>$C=C$<$ \ $\rightarrow$ ($\downarrow$X : \ ) $\rightarrow$ \ $^{\rm H}>$C$_-^{\rm X}$C$<$  \ $\rightarrow$ (h$\nu^\star$ : $^\uparrow$XH) $\rightarrow$ \ \ \ \ $>$C=C$<$ \ \ $\rightarrow$ \ \ \ \ (H : \ ) \ \ \ \ \ \ $\rightarrow$ $^{\rm H}>$C=C$<$ \\[0.1cm]
\hspace*{0.3cm} $>$C=C$<$ \ $\rightarrow$ ($\downarrow$X : \ ) $\rightarrow$ \ \ \ \ \ $>$C$_-^{\rm X}$C$<$ \ $\rightarrow$ \ ($\downarrow$H : \ ) \ \ \ \ \ \ \ $\rightarrow$ \ \ $>$C$_-^{\rm X^{\rm H}}$C$<$ \ $\rightarrow$ \ (h$\nu^\star$ : $^\uparrow$XH) \ $\rightarrow$ \ $>$C=C$<$ \\[0.2cm] 
where we adopt the descriptor (reactant : product): with $h\nu^\star$ being reaction-driving UV photons.
The likely pathways for the formation of nano-particle surface epoxide and aziridine groups and their subsequent photolysis/surface catalysis routes to OH and NH radical formation, and molecular hydrogen formation via an aziridine, are shown in more detail in Figs. \ref{fig_epoxide_OH} and \ref{fig_aziridine_NH}. 

The rate-determining step in the formation of OH and NH via the above epoxide and aziridine pathways, respectively, is the X atom (O or N) chemisorption rate onto nano-particles, which is given  by:
\begin{align}\label{eq_H_int}  
R_{\rm X} = n_{\rm H} \ X_{\rm X} \ \pi a_{\rm np}^2 \ v_{\rm X} \ s_{\rm X},
\end{align}
where $n_{\rm H}$ is the H atom density,   
$X_{\rm X}$ is the gas phase abundance of atom X relative to hydrogen,  
$v_{\rm X}$ its thermal speed and $s_{\rm X}$ its sticking efficiency. Adopting, $a = 0.4$\,nm a-C(:H) nano-particles in the atomic ISM,\cite{2013A&A...558A..62J}  
$n_{\rm H} = 30$\,cm$^{-3}$, $X_{\rm O} = 500$\,ppm, $X_{\rm N} = 100$\,ppm, $s_{\rm O} = s_{\rm N} = 1$ and $v_{\rm O} = v_{\rm N} \simeq 3 \times 10^4$\,cm\,s$^{-1}$, the formation rate for OH is then $R_{\rm OH} = 2.3 \times 10^{-12}$\,s$^{-1}$.   
For NH the formation rate will probably be at least a factor of five less than that for OH, given the nitrogen relative abundance and all other parameters equal. However, nitrogen appears to have an affinity for a-C(:H) materials and appears to promote sp$^2$ cluster formation.\cite{1997PhRvB..5513020L,2001JAP....89.7924H,2004PhRvB..69d5410Y,2006TSF...515.1597P}
Hence, once incorporated into a-C(:H) nitrogen atoms may be harder to remove by FUV photolysis than epoxide-bonded oxygen atoms, and so NH formation by this route is likely to be intrinsically less efficient than OH formation. 
The derived OH and NH formation rates will likely hold for a-C(:H) nano-particles ($a \lesssim 1$\,nm) in the diffuse/translucent ISM with low extinction, {\it i.e.}, $A_{\rm V} \lesssim 0.5$\,mag., where the UV photon flux is greater than the O and N atom fluxes. 

In the ISM surface-epoxide-formed OH will be photo-dissociated, at a rate $R_{\rm diss.} \simeq 2 \times 10^{-10}$\,s$^{-1}$,\cite{1985ApJ...291..202N} once released into the gas phase in low extinction regions ($A_{\rm V} \lesssim 1$\,mag.).  
From the reaction kinetics we have that 
\begin{align}\label{eq_H_int}
\frac{dn{\rm [OH]}}{dt} = -R_{\rm diss.} \ n_{\rm OH} \ + \  R_{\rm OH} \ n_{\rm H} \  X_{\rm np}, 
\end{align}
where $X_{\rm np}$ is the nano-particle abundance relative to hydrogen. 
For the almost mono-modal a-C(:H) nano-particle size distribution in the low density atomic ISM $X_{\rm np} \simeq 5 \times 10^{-6}$ for $a = 0.4$\,nm ($\simeq 150$\,ppm of carbon out of a total dust budget carbon requirement of $\simeq 200$\,ppm).\cite{2013A&A...558A..62J}  
The above-derived OH formation rate, $R_{\rm OH} = 2.3 \times 10^{-12}$\,cm$^{-3}$\,s$^{-1}$ along with $X_{\rm O} = 500 \times 10^{-6}$, then implies an OH relative abundance in these regions given by $n_{\rm OH}/n_{\rm H} = (R_{\rm OH} \, X_{\rm np})/R_{\rm diss.} = 5.7 \times 10^{-8}$. 
For the formation of NH using the same photo-dissociation rate as for oxygen, but assuming that some fraction, $f_{\rm N,np}$, of the nitrogen atoms become trapped into aromatic moiety peripheral sites in nano-particles, we find $n_{\rm NH}/n_{\rm H} \lesssim (1-f_{\rm N,np}) \times 10^{-9}$ and suggest that the NH abundance will be at least an order of magnitude less than that of OH in a given region. 

In the tenuous or translucent ISM OH appears to be rather widespread and to follow atomic hydrogen (HI), both spatially and in velocity extent, rather than CO.\cite{2010MNRAS.407.2645B,2012AJ....143...97A} OH appears to correlate with the dust long wavelength emission at $100\,\mu$m and to increase in abundance with the visual extinction above a threshold of $A_{\rm V} \gtrsim 0.5$\,mag. ($\equiv N_{\rm H} \gtrsim (2.2-2.9) \times 10^{20}$\,cm$^{-2}$; similar to the observed threshold for H$_2$ formation, {\it i.e.}, $N_{\rm H} \gtrsim (2.9-3.6) \times 10^{20}$\,cm$^{-2}$).\cite{2010MNRAS.407.2645B}  The   abundance of OH relative to atomic HI observed in the cirrus clouds in the North Celestial Loop is $N_{\rm OH}/N_{\rm HI} = (6.5-7.4) \times 10^{-8}$.\cite{2010MNRAS.407.2645B} 
The above-estimated photolysis-driven OH relative abundance ($5.7 \times 10^{-8}$) for the low-density ISM therefore appear to be in very good agreement with the observed OH abundances. 
The fundamental process underlying this proposed photolytic OH formation route is compatible with a similar mechanism for the photolysis-driven formation of H$_2$ in moderately-excited photo-dissociation regions.\cite{2015A&A...581A..92J}

\begin{figure}[!h]
\centering\includegraphics[width=4.0in]{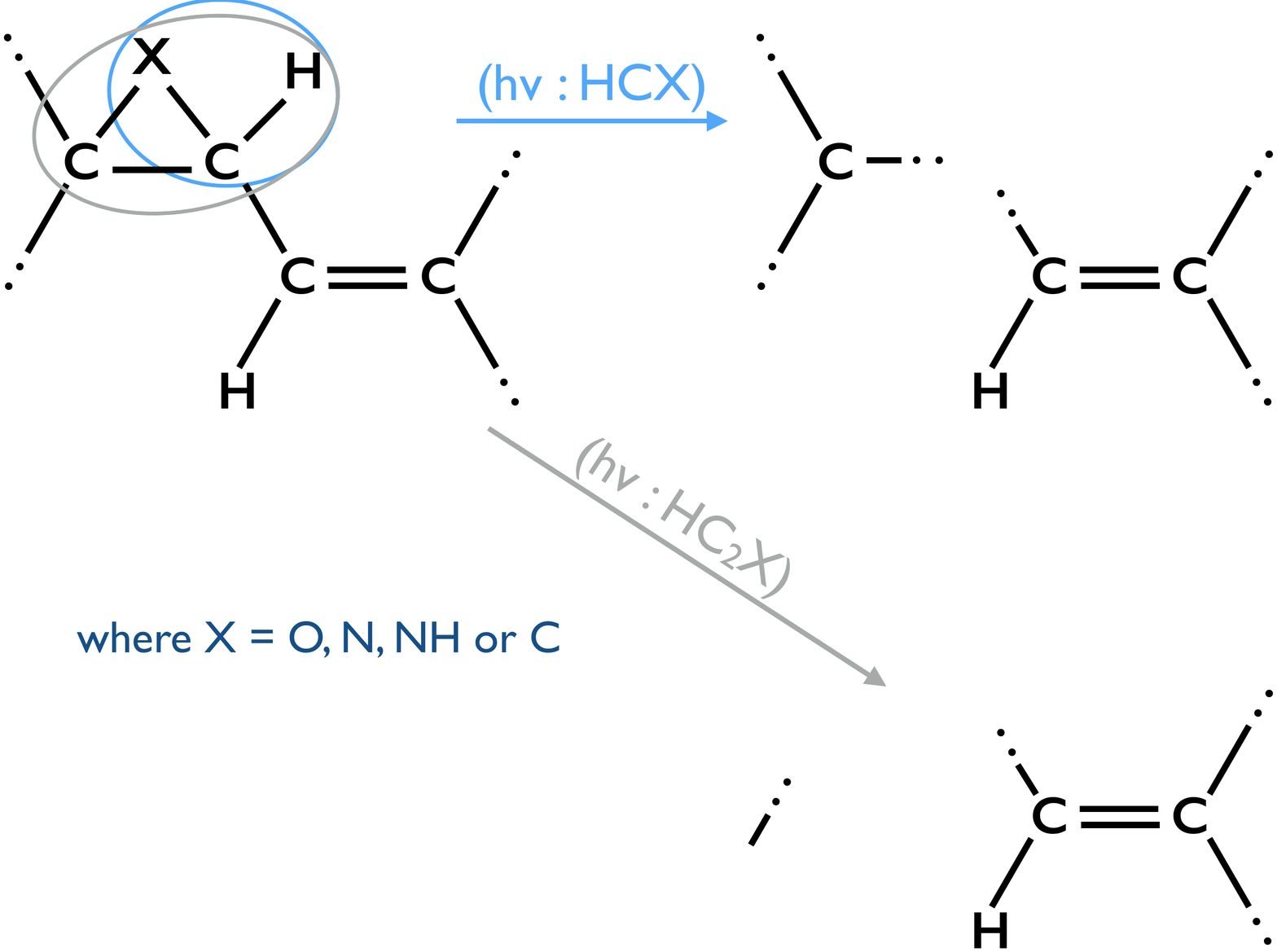}
\caption{An epoxide/aziridine FUV photolysis-driven pathway to nano-particle erosion in the tenuous ISM. In this case the product radicals could include: HCO, HC$_2$O, HCN, HNC and HCNH and perhaps also HC$_2$NH, l-C$_3$H and c-C$_3$H if C$^+$ inserts into olefinic C$=$C bonds to form a cyclopropene ring, somewhat analogous to epoxide and aziridine formation.}
\label{fig_np_dest}
\end{figure}

In the above it was assumed that the heavy atom incorporation into nano-particles and the resultant catalytic formation of OH and NH radicals is recyclable, {\rm i.e.},  there is no inherent destruction of the grain structure. However, and given that things seem to be rather finely balanced in the tenuous ISM, it is possible that the scales could be tipped towards FUV photolysis leading to the loss of radical species with two heavy atoms, therefore resulting in nano-particle erosion. This type of a-C(:H) erosional pathway is illustrated in Fig.~\ref{fig_np_dest}, which indicates that some interesting radical species could be liberated into the gas where they would be ionised and perhaps also dehydrogenated.

\section{Nitrogen}
\label{sect_N}

\begin{figure}[!h]
\centering\includegraphics[width=4.0in]{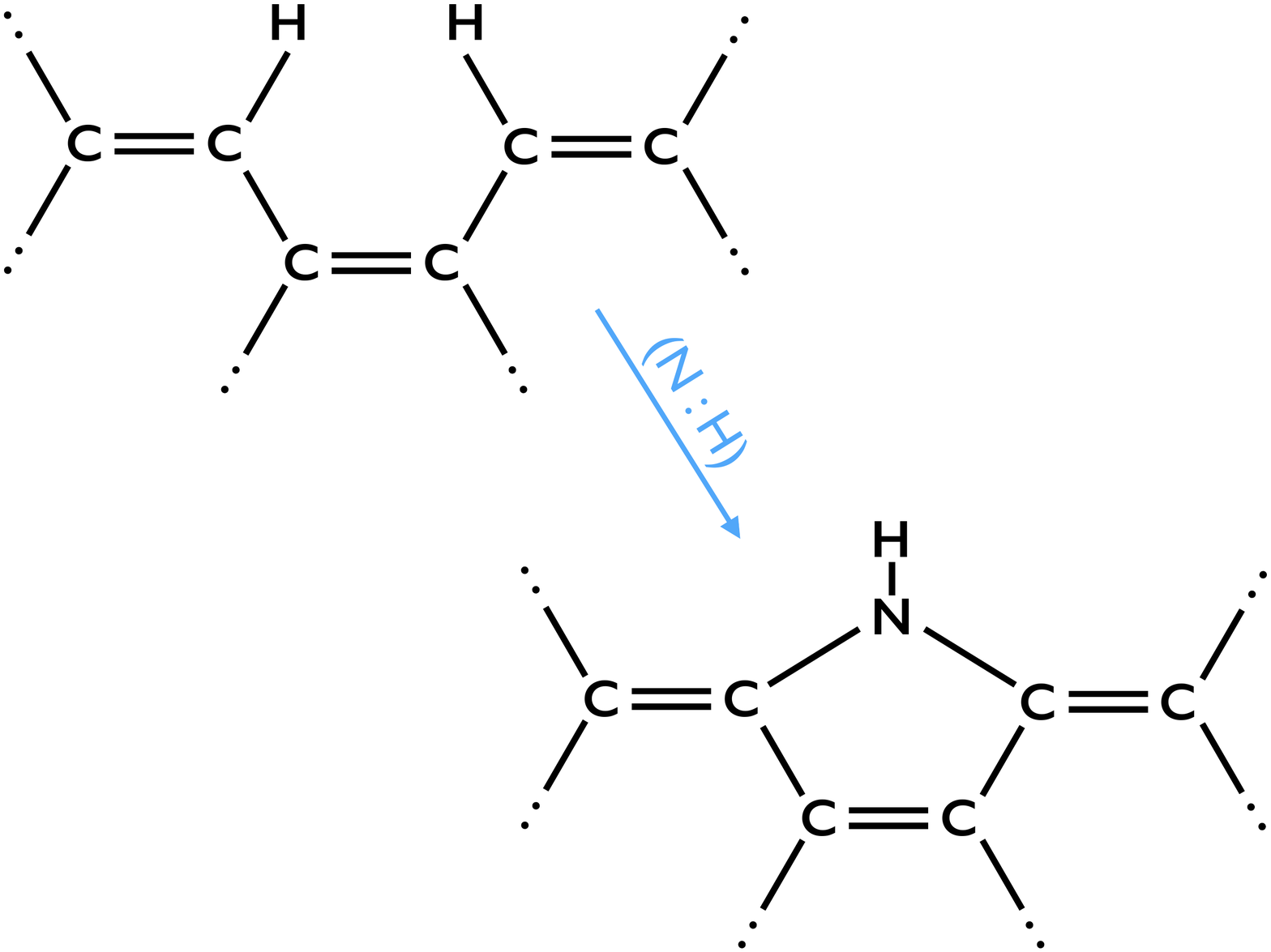}
\caption{A route to N-containing, hetero-atom, five-fold rings in poly-olefinics and on the periphery of aromatic moieties.}
\label{fig_Npentagons}
\end{figure}

\begin{figure}[!h]
\centering\includegraphics[width=4.0in]{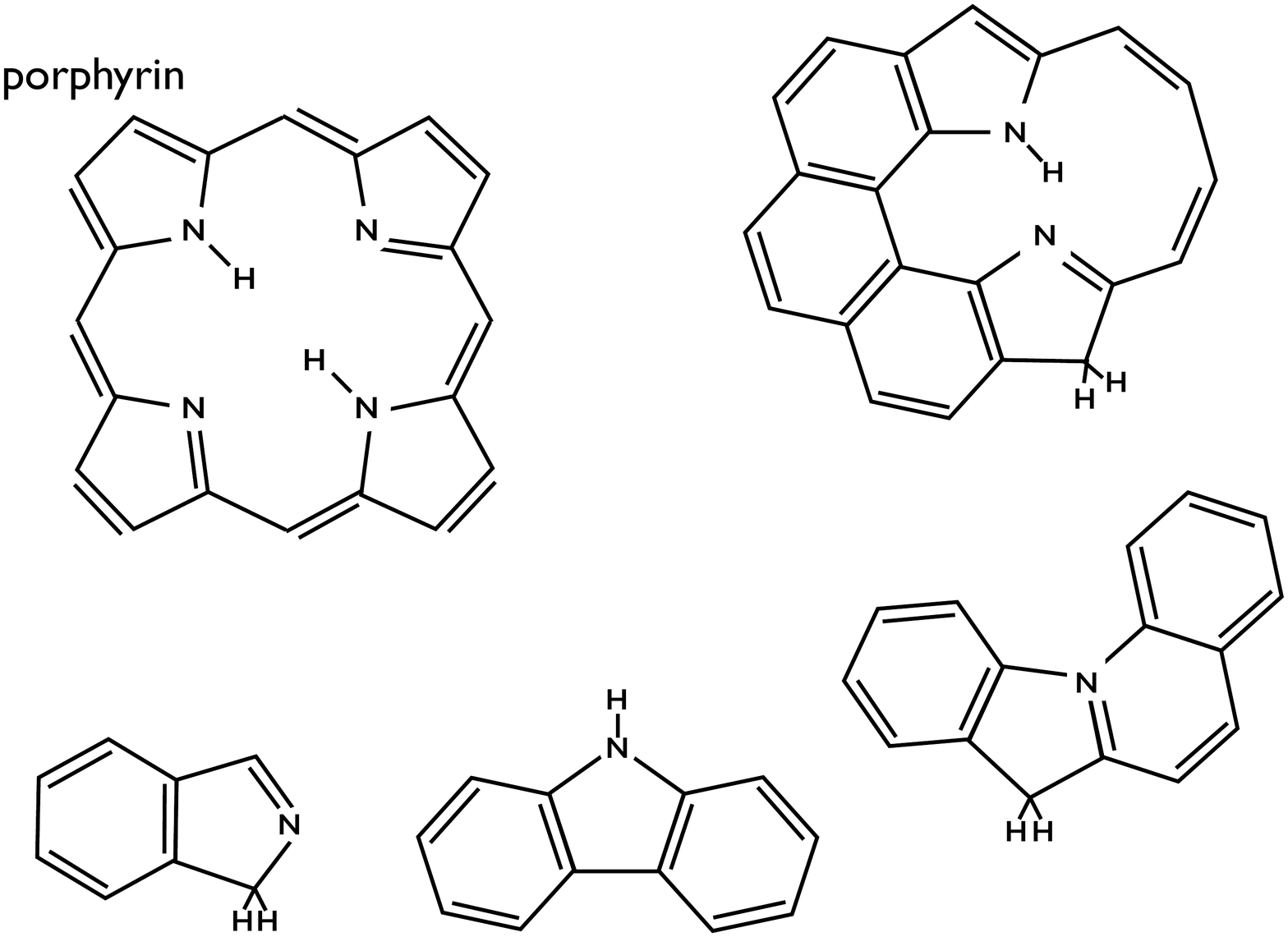}
\caption{A pophyrin structure (upper left) and possible porphyrin precursor moieties typical of asphaltene-type structures and aromatic domains within N-doped a-C(:H) nano-particles (increasing complexity from bottom left to upper right).}
\label{fig_porphyrin}
\end{figure}

Given the affinity of nitrogen for incorporation into a-C(:H) materials\cite{1997PhRvB..5513020L,2001JAP....89.7924H,2004PhRvB..69d5410Y,2006TSF...515.1597P} a possible key role of nitrogen hetero-atoms within interstellar carbonaceous grains has been considered within the astrophysical context\cite{2013A&A...555A..39J,2014P&SS..100...26J,ANT_RSOS_topdown} and from this work it appears that a doping level of only the order of one percent or so, with respect to carbon, would yield interesting effects.\cite{2013A&A...555A..39J,2014P&SS..100...26J} 
Indeed, such a doping effect has been proposed as being an important element in our search for an explanation for the nature of the diffuse interstellar band (DIB) carriers\cite{ANT_RSOS_topdown}. 

It is likely that N-doped a-C(:H) cage structures, fullerene-type structures and cage-trapped metals could provide a route to N-containing hydrocarbon clusters analogous to porphyrin-type species, where the nitrogen atoms become preferentially trapped in five-fold rings as appears to be a common feature of asphaltene-type aromatic moieties.\cite{JACS_2015_Asphaltenes,ANT_RSOS_topdown}   
Indeed, a possible route to peripheral, nitrogen-containing, five-fold rings on olefinics and aromatics is shown in Fig.~\ref{fig_Npentagons}. Further, these types of structures, as schematically shown in Fig.~\ref{fig_porphyrin}, could provide a stepping stone to the formation of the precursors of porphyrin-type moieties. Porphyrins are key elements in the structure of chlorophyll and heme when associated with centrally-coordinated magnesium and iron  atoms, respectively. Interestingly, porphyrin structures exhibit a strong blue absorption band in the $\sim 430-450$\,nm wavelength region, which is intriguingly close to the strongest DIB at 443\,nm. Indeed, porphyrins would appear to have many bands in common with the DIBs.\cite{Johnson_porphyrin_and_DIBs}

Chlorophyll molecules (C$_{35}$H$_{28-30}$O$_{5}$N$_{4}$Mg, type c1 and c2; C$_{54-55}$H$_{70-72}$O$_{5-6}$N$_{4}$Mg, type a, b, d and e) are basically aromatic-type clusters with a magnesium atom at their centre and a long alkane chain about 15 C atoms long, with ether bridges and carbonyl groups at it's base. 
Likewise within h\ae moglobin the heme structure (C$_{34}$H$_{32-36}$O$_{4}$N$_{4}$S$_{0-2}$, type b or c; C$_{49}$H$_{55-56}$O$_{4-6}$N$_{4}$Fe, type a or o) has a porphyrin structure with a central iron atom and carboxylic acid-terminated alkyl side groups. In chlorophyll the alkyl chain's primary role appears to be to allow the molecule to incorporate within hydrophobic environments but, other than this key requirement, other possible roles for it do not seem to have been given much consideration. The long alkane chain, with carbonyl and ether groups at its base, is indeed a rather curious beast and it appears that such a structure would provide an ideal energy absorber to enable the molecule to dissipate the energy associated with photon absorption through vibrational excitation. Hence, the alkyl chain may have evolved into an energy absorber that stabilises the molecule against excitation in the same way that the aliphatic/olefinic bridges and/or tails within the contiguous networks of a-C(:H) may play an energy-absorbing role in interstellar nano-particles (see Section 3 subsections~\ref{sect_asphaltenes} and \ref{sect_jelly}).
It is interesting that long alkyl side groups, similar in length to those of chlorophyll molecules, are a rather common feature of the analysed asphaltenes extracted from coal and oil.\cite{JACS_2015_Asphaltenes} 
Thus, the chlorophyll structure appears to bear a generically rather striking resemblance to the structures observed within asphaltene moieties. 
This is perhaps not too surprising given that asphaltenes originate from the decay of organic plant matter, a process that has occurred over geological time-scales.

\section{Natural selection}
\label{sect_natsel}

Interstellar a-C(:H) nano-particles with up to a few hundred heavy atoms can probably exist in tens of millions of different  configurational forms.\cite{ANT_RSOS_topdown} 
It is interesting that the key porphyrin-based structures within the chlorophyll and heme molecules contain only of the order of $\sim 40-70$ and $\sim 40-60$ heavy atoms, respectively, which is considerably fewer atoms than a typical interstellar nano-particle. 
Further, and as mentioned above, such generic types of structures are a feature of the naturally-occurring asphaltene moieties. 

Thus, extra-solar system nano-particles incorporated into larger grains and small solar system bodies in the pre-solar nebula, raining down through a planetary atmosphere onto the surface, would provide a very rich resource from which to inorganically select the more stable and later-to-be-useful species. A selection process would likely occur through photo-, thermal- and aqueous-processes operating over geological time-scales. Thus, viable solar energy absorbing structures ({\it a.k.a}, chlorophyll) and trappers and transporters of useful molecules within and across particle boundaries ({\it a.k.a}, O$_2$, CO transport by h\ae moglobin) could have been selected over long time-scales to provide a source of pre-pre-cursor, pre-biotic molecules that would act as complex chemistry drivers on the primitive Earth. These reactions would eventually lead this chemistry along a route to the first simple self-replicating biotic species. Thus, while life may not have come from space it was likely seeded with a wide variety of species that would inevitably lead to the appearance of life on the Earth. 

Thus, even before natural selection could operate on living organisms, the stage for it's operation at the molecular level may already have been set long before any self-replicating structures or species appeared on Earth.

\subsection*{Implications}
\label{sect_implications}

A route towards the pre-cursor\footnote{Pre-cursor in this sense is taken to mean the chemical species that pre-date biotic molecules by a few major steps, these being: biotic molecules derived from pre-biotic molecules (inorganic species of relevance to biology but not biologically active), which themselves evolved from extraneous matter (extra-planetary system solid bodies).} seeding of later-formed key (pre-)biotic molecules appears as though it could have been influenced by the intrinsic nature of pre-exisitng interstellar carbonaceous grains polluted with nitrogen, oxygen and other less-abundant hetero-atoms. 
Once these interstellar grains are transported to a planetary surface they at last find themselves in a protective environment shielded from the effects of harmful radiation in the ISM. On a planetary surface these grains will provide a highly diverse and valuable feedstock from which the local conditions would down-select the most favourably-stable structures for that given environment. 
This would most likely occur in the near-surface sediments of an aqueous environment, where the effects of the proto-stellar radiation field are suitably attenuated but sufficiently intense to provide an energy source for chemical reactions. 
Among these reactions simple ion-trapping by solids from the overlying solution (primeval soup?) could lead to the incorporation of metal cations within "selected" pre-cursor inorganics (aromatic-type moieties) of extra-planetary origin. 

It then appears that the pre-cursor species could provide a natural but slow route to pre-biotic species including pre-porphyrins that would over the millennia evolve into interesting chlorophyll-like and heme-like species, and a whole host of other simpler and more complex molecules and structures.

Thus, natural selection is probably also an important mechanism, that operates at the molecular level in a primitive planetary surface environment, to weed out the less-than-useful chemical species and make the best use of stable species with "valuable" properties. 
A major implication from this line of reasoning is that, given the same source materials of interstellar origin, the chemical and biological evolution leading to life probably follows similar pathways in similar environments, leading to essentially the same underlying "chemistry of life" everywhere.

\section{Conclusions} 
\label{sect_conclusions}

The chemistry and physics of (interstellar) nano-particles is very particular, in that important size-dependent effects have to be taken into account when modelling the effects of the environment on nano-particles (evolution through erosion, mantling, coagulation, \ldots) and their return effects on that environment (photo-electric heating of the gas, formation of molecular hydrogen and small radicals by surface reactions, \ldots). Thus, a full understanding of their properties will be essential in advancing our understanding of the evolution of the low density ISM. 

In no small part, interstellar chemistry in transition regions, the clouds in the stages of evolution between the most diffuse  clouds and molecular clouds, is probably driven by the active surface chemistry of nascent, carbonaceous, a-C(:H), nano-particles. The catalytic FUV photon-driven formation of OH, NH and other small radicals and molecules ({\it e.g.}, HCO, HC$_2$O, HCN, HNC, HCNH, HC$_2$NH, l-C$_3$H and c-C$_3$H, \ldots) on and within nano-particles may therefore be the initiator of the chemistry in the most tenuous regions of the ISM. Further, it is also likely that interstellar nano-particles will retain their nascence whether in a free-flying state or coagulated into aggregates, provided that they are not submerged under an "inert" molecular ice mantle.  

It  appears likely that nitrogen plays an important role within interstellar grains and may, even at low doping levels ($\simeq 1$\%) be a key element in explaining the origin of the diffuse interstellar bands. Having a preference for incorporation within the five-fold hetero-cycles associated with aromatic-rich moieties, nitrogen is a key ingredient in many interesting molecules ({\it e.g.}, porphyrins, chlorophyll, heme, \ldots) where it is present at the level of only a few atomic percent. The importance of nitrogen in dust, and particularly in nano-particles, has therefore most likely been overlooked because the low levels of depletion required to yield interesting effects. Small nitrogen depletions are hard to detect and, in any event, the direct association of nitrogen with any observable measure of dust in the low density ISM has yet to be made. 

It is likely that planetary surfaces will be seeded with pre-cursor species from a universal source (interstellar and interplanetary medium dust,  nano-particles in particular). 
These particles will provide a rich chemical variety of nanometre-sized structures, principal among them are likely to be analogues of the aromatic-rich moieties extracted from terrestrial coals and oils (asphaltenes). 
The principal effect of the planetary surface processing of this rich resource, most likely in shallow aqueous bodies, would be to naturally down-select the most stable species that could, over geological time-scales, evolve into useful pre-biotic molecules.  
If these seed species are a significant link in the chain of events leading to life, then the origin of life will most likely follow the same chemical pathways and be rather similarly constructed everywhere in the Universe, wherever and whenever it arises. 
Whether intelligence arises as a natural consequence of this process remains a matter for debate.

%


\footnotesize{
\bibliography{Ant_bibliography} 
\bibliographystyle{rsc} 
}

\end{document}